\documentclass[intlimits,twoside,a4paper]{article}

\usepackage{amsmath,amssymb}
\usepackage{graphicx}

\usepackage[T2A]{fontenc}
\usepackage[cp1251]{inputenc}

\usepackage[eqsecnum]{cmpj2}

\issue{2016}{19}{3}{33702}
\doinumber{10.5488/CMP.19.33702}

\title[Method of computation in FQHE]%
{Method of computation of energies in the fractional quantum Hall effect regime}
\author[M.A.~Ammar, Z.~Bentalha, S.~Bekhechi]{M.A.~Ammar\refaddr{label1},
        Z.~Bentalha\refaddr{label2},
        S.~Bekhechi\refaddr{label2} }
\addresses{
\addr{label1} Environment Department, University of Medea, 26000 Media, Algeria
\addr{label2} Theoretical Physics Laboratory, University of Tlemcen, B.P. 230, 13000 Tlemcen, Algeria
}

\authorcopyright{M.A.~Ammar, Z.~Bentalha, S.~Bekhechi, 2016}
\date{Received March 12, 2016, in final form April 18, 2016}

\begin{document}

\maketitle

\begin{abstract}
In a previous work, we reported exact results of energies of the ground state in the fractional quantum Hall effect (FQHE) regime for systems with up to $N_{\text{e}} = 6$ electrons at the filling factor $\nu = 1/3$ by using the method of complex polar coordinates. In this work, we display interesting computational details of the previous calculation and extend the calculation to $N_{\text{e}} = 7$ electrons at $\nu = 1/3$. Moreover, similar exact results are derived at the filling $\nu = 1/5$ for systems with up to $N_{\text{e}} = 6$ electrons. The results that we obtained by analytical calculation are in good agreement with their analogues ones derived by the method of Monte Carlo in a precedent work.
\pacs 73.43.-f, 73.43.Cd, 71.10.Ca, 02.70.Wz
\end{abstract}
\section{Introduction}
The discovery of the fractional quantum Hall effect (FQHE)~\cite{R1} was the beginning of a big revolution in the field of condensed matter. Since then, new concepts of matter state have been raised such as the incompressible quantum fluid~\cite{R2}, composite fermions~\cite{R3,R4,R5,R6}, composite bosons~\cite{R7} and anyons~\cite{R8,R9}, all emanating from elegant theories with sophisticated mathematics. Nowadays, there are two world wide accepted theories in the field of FQHE, the theory of Laughlin~\cite{R2} and the theory of Jain~\cite{R3,R4,R5,R6}. The former describes the ground state as an incompressible quantum fluid which successfully clarified the nature of states at the filling factors $\nu = 1/3, 1/5, 1/7, \ldots\,$. The latter (theory) is built upon the concept of composite fermions that are topological entities caricaturing the idea of  electrons embracing a number (even) of quantized vortices and gives satisfactory results regarding the $\nu=p/(2mp+1)$ states, integer $m$ and~$p$. In Laughlin theory, the incompressible quantum fluid consists of strongly correlated electrons interacting with a strong magnetic field whereas in Jain theory it consists of weakly correlated composite fermions interacting with a reduced magnetic field. The aim of determining ground state energies for FQHE electron systems has been the object of many investigations with various computational methods such as exact diagonalization~\cite{R10,R11,R12,R13,R14}, density matrix renormalization group~\cite{R15} or Monte Carlo simulations~\cite{R16,R17,R18}. However, it is worth notifying that at the level of quasiparticle state, certain discrepancy is observed between the results of~\cite{R19} using spherical geometry and the results of~\cite{R20} using disk geometry, while the authors of~\cite{R20} have found the reason of the discrepancy to be unclear. Also, all these methods are numerical and one may wonder whether it is possible to perform analytical methods that would serve as reliable comparison instruments even for small systems of electrons. The most pronounced analytical method in this line of research was given by the author of~\cite{R22}, where ordinary polar coordinates (including ordinary Jacobi coordinates) are employed but technical calculational difficulties arise for systems with $N_{\text{e}}>4$ electrons, and one can see an explicit dependence of the integrands upon the angles of the particles (see equation (21) in~\cite{R22}). This apparent difficulty can be overcome by using complex polar coordinates, being the main contribution of~\cite{R21}. By the way, it should be notified that ordinary polar coordinates are also used in the original paper by Laughlin~\cite{R2} within a Monte Carlo study. Regarding the work~\cite{R21}, we developed an analytical method based on complex polar coordinates and explicitly calculated the energies of the ground state in the FQHE regime for systems with up to six electrons at the filling $\nu = 1/3$. The use of polar coordinates in a complex form has been the key tool which greatly simplifies the calculation of some complicated expressions involving integrals over many variables in~\cite{R21}. The aim of this work is to show all the necessary computational steps and details underlying the analytical method of~\cite{R21} so as to make its application possible in other areas of condensed matter physics, especially in 2D~Coulomb systems such as 2D Dyson gas wherein expressions involving integrals over many variables are encountered such as the partition function or the mean energy. Thus, we obtain new exact analytical results concerning the energy of the ground state for systems with $N_{\text{e}} = 7$ electrons at the filling $\nu = 1/3$, and $N_{\text{e}} = 5, 6$ electrons at the filling $\nu = 1/5$.

The paper is organized as follows. In section~\ref{sec:1}, the theoretical background is presented. In section~\ref{sec:2}, the electron-electron interaction energy is calculated. In section~\ref{sec:3}, the method of computation of the electron-background interaction energy is shown. In section~\ref{sec:4}, we give the results of our calculus. Section~\ref{sec:6} is devoted to the conclusion.
\section{Theoretical background}
\label{sec:1}
We consider $N_{\text{e}} ( > 2)$ electrons of charge ($-e_{0}$) embedded in a uniform neutralizing background disk of positive charge $N_{\text{e}}\,e_{0}$ and area $S_{N_{\text{e}}}=\pi R_{N_{\text{e}}}^{2}$, $R_{N_{\text{e}}}$ is the radius of the disk. We also assume that the disk is a part of the $XY$ plane subjected to a strong uniform magnetic field, in the $z$ direction, ${\bf{B}}=B\,\bf{e_{z}}$. The physics of the FQH fluid is then governed by a full interaction potential
\begin{equation}
V=V_{\text{ee}}+V_{\text{eb}}+V_{\text{bb}}
\label{E1}
\end{equation}
with $V_{\text{ee}}, V_{\text{eb}}$ and $V_{\text{bb}}$ denoting the electron-electron, electron-background and the background-back\-ground interaction potentials, respectively. Their corresponding expressions are given by
\begin{eqnarray}
V_{\text{ee}}=\sum_{i<j}^{N}\frac{e_{0}^{2}}{\mid {\bf{r}}_{i}- {\bf{r}}_{j} \mid}\,,
\label{E2}
\end{eqnarray}
\begin{eqnarray}
V_{\text{eb}}=-\rho\sum_{i=1}^{N}\int_{S_{N}}\rd^{2}r\frac{e_{0}^{2}}{\mid {\bf{r_{i}}}- {\bf{r}} \mid}\,,
\label{E3}
\end{eqnarray}
and
\begin{eqnarray}
V_{\text{bb}}=\frac{\rho^{2}}{2}\int_{S_{N}} \rd^{2}r \int_{S_{N}} \rd^{2}r^{\prime}\frac{e_{0}^{2}}{\mid {\bf{r}}- {\bf{r^{\prime}}} \mid}\,,
\label{E4}
\end{eqnarray}
where ${\bf{r}}_{i}$ (or ${\bf{r}}_{j}$) indicate the electron vector position while ${\bf{r}}$ and ${\bf{r^{\prime}}}$ are background coordinates. $S_{N_{\text{e}}}(B)$ is the area of the disk and $\rho(B)$ is the density of the system (the number of electrons per unit area) that can also be defined by
\begin{equation}
\rho=\frac{\nu}{2\pi l_{0}^{2}}\,,
\label{E5}
\end{equation}
where $l_{0}(B)=\sqrt{\hbar\,c/(e_{0} B)}$ is the magnetic length, $c$ is the speed of light, $B$ is the magnetic field strength, and $\nu=1/m$ is the filling factor, $m=3,5,\ldots\,$. The background-background interaction potential can be classically calculated without using the wave function of the electron system. Its value is simply determined by calculating the elementary defined integral (\ref{E4}) and is given by~\cite{R22}
\begin{equation}
V_{\text{bb}}=\frac{8e_{0}^{2}}{3\pi}\frac{N_{\text{e}}}{R_{N_{\text{e}}}}
\label{E6}
\end{equation}
with $R_{N_{\text{e}}}=\sqrt{2N_{\text{e}}m}\,l_{0}$. It remains to calculate the energies corresponding to the electron-electron and the electron-background potentials which depend on the nature of the wave function characterizing the system of electron.

For a given wave function $\Psi({\bf{r}}_{1},{\bf{r}}_{2},\ldots ,{\bf{r}}_{N_{\text{e}}})$, these energies are determined using the following formulae
\begin{align}
\langle V_{\text{ee}}\rangle & =  \frac{\langle\Psi\vert V_{\text{ee}}\vert\Psi \rangle}{\langle\Psi\vert \Psi \rangle}\,,\label{E7} \\
\langle V_{\text{eb}}\rangle & =  \frac{\langle\Psi\vert V_{\text{eb}}\vert\Psi \rangle}{\langle \Psi \vert \Psi \rangle}\,.
\label{E8}
\end{align}
 In an explicit manner, we have
\begin{align}
\langle\Psi\vert V_{\text{ee}}\vert\Psi \rangle  &=  \frac{N_{\text{e}}(N_{\text{e}}-1)}{2}\int \rd^{2}r_{1}\ldots \rd^{2}r_{N_{\text{e}}} \frac{e_{0}^{2}}{\mid {\bf{r}}_{1}- {\bf{r}}_{2} \mid} \mid \Psi({\bf{r}}_{1},\ldots ,{\bf{r}}_{N_{\text{e}}})\mid^{2}, \label{E9}\\
\langle\Psi\vert V_{\text{eb}}\vert\Psi \rangle  &=  -\rho N_{\text{e}} \int \rd^{2}r_{1}\ldots \rd^{2}r_{N_{\text{e}}}\mid \Psi({\bf{r}}_{1},\ldots ,{\bf{r}}_{N_{\text{e}}})\mid^{2} \int_{S_{N_{\text{e}}}}\rd^{2}r \frac{e_{0}^{2}}{\mid {\bf{r}}_{1}- {\bf{r}} \mid}\,,  \label{E10}\\
\langle \Psi \vert \Psi \rangle  &=  \int \rd^{2}r_{1}\ldots  \rd^{2}r_{N_{\text{e}}} \mid \Psi({\bf{r}}_{1},\ldots , {\bf{r}}_{N_{\text{e}}})\mid^{2}
\label{E11}
\end{align}
with (\cite{R23,R24})
\begin{equation}
\int_{S_{N_{\text{e}}}}\rd^{2}r \frac{1}{\mid {\bf{r}}_{1}- {\bf{r}}\mid}=2\pi\, R_{N_{\text{e}}} \int_{0}^{\infty} \frac{\rd q}{q}J_{1}(q)J_{0}\left(\frac{q}{R_{N}}r_{1}\right),
\label{E12}
\end{equation}
where $J_{n}(x)$ are $n$-th order Bessel functions.

However, as shown in~\cite{R21}, the best way of making numerous simplifications in subsequent calculations amounts to replacing real polar coordinates by complex polar coordinates. Thus, let us apply the change ${\bf{r}}_{k} \longrightarrow z_{k}=(x_{k}+\ri y_{k}=r_{k}e^{\ri\varphi_{k}})_{k=1,\ldots ,N}$ to localize the electrons.
\section{The $\langle V_{\text{ee}}\rangle$ calculation}
\label{sec:2}
Now, for a demonstrative calculation, we will focus on the case of $N_{\text{e}}=4$ electrons and $m=3$. Let $\Psi$ be the wave function of Laughlin for $N_{\text{e}}=4$ electrons and $m=3$
\begin{equation}
\Psi(4)=P(4)\:\exp\left(-\sum_{k}\frac{\mid z_{k}\mid^{2}}{4l_{0}^{2}}\right),
\label{E13}
\end{equation}
where $P(4)$ is the Jastrow part of the wave function that is given by
\begin{equation}
P(4)= (z_{1}-z_{2})^{3}(z_{1}-z_{3})^{3}(z_{1}-z_{4})^{3}(z_{2}-z_{3})^{3}(z_{2}-z_{4})^{3}(z_{3}-z_{4})^{3}.
\label{E14}
\end{equation}
In complex coordinates, the expressions of equation (\ref{E9}) and equation (\ref{E11}) transform into
\begin{align}
\langle\Psi\vert V_{\text{ee}}\vert\Psi \rangle  &=  \frac{N_{\text{e}}(N_{\text{e}}-1)}{2}\int \rd^{2}z_{1}\ldots \rd^{2}z_{N_{\text{e}}} \frac{e_{0}^{2}}{\mid z_{1}- z_{2} \mid} \mid \Psi(z_{1},\ldots ,z_{N_{\text{e}}})\mid^{2},  \label{E15}\\
\langle \Psi \vert \Psi \rangle  &=  \int \rd^{2}z_{1}\ldots  \rd^{2}z_{N_{\text{e}}} \mid \Psi(z_{1},\ldots , z_{N_{\text{e}}})\mid^{2}.
\label{E16}
\end{align}
Now, we should perform a Jacobi transformation with complex coordinates instead of real coordinates \cite{R21} so as to get rid of the term $\mid z_{1}-z_{2}\mid$ in the denominator of the integrand of the expression~(\ref{E15}), which is done using the following:
\begin{align}
Z_{1} &=  z_{1}-z_{2},  \\
Z_{2} &=  \frac{z_{1}}{2} + \frac{z_{2}}{2} - z_{3}, \\
Z_{3} &=  \frac{z_{1}}{3} + \frac{z_{2}}{3} + \frac{z_{3}}{3} - z_{4}, \\
Z_{4} &=  \frac{z_{1}}{4} + \frac{z_{2}}{4} + \frac{z_{3}}{4} + \frac{z_{4}}{4}.
\label{E17}
\end{align}
Then, the inter-particle coordinates $(z_{i} - z_{j})$ can be written in terms of Jacobi coordinates as follows:
\begin{align}
z_{1}-z_{2} &=  Z_{1},  \\
z_{1}-z_{3} &=  Z_{2} + \frac{Z_{1}}{2},  \\
z_{1}-z_{4} &=  Z_{3} + \frac{Z_{2}}{3} + \frac{Z_{1}}{2},  \\
z_{2}-z_{3} &=  Z_{2} - \frac{Z_{1}}{2}, \\
z_{2}-z_{4} &=  Z_{3} + \frac{Z_{2}}{3} - \frac{Z_{1}}{2},  \\
z_{3}-z_{4} &=  Z_{3} - 2\frac{Z_{2}}{3}.
\label{E18}
\end{align}
Now, to write $P(4)$ in terms of the $Z_{i}$ Jacobi coordinates, we just recast (\ref{E18}) into (\ref{E14}), then there holds the polynomial
\begin{eqnarray}
P_{J}(4) = Z_{1}^{3}\left(Z_{2} + \frac{Z_{1}}{2}\right)^{3}\left(Z_{3} + \frac{Z_{2}}{3} + \frac{Z_{1}}{2}\right)^{3}
\left(Z_{3} + \frac{Z_{2}}{3} + \frac{Z_{1}}{2}\right)^{3}\left( Z_{2} - \frac{Z_{1}}{2}\right)^{3}
\left(Z_{3} + \frac{Z_{2}}{3} - \frac{Z_{1}}{2}\right)^{3}\left(Z_{3} - 2\frac{Z_{2}}{3}\right)^{3}.
\label{E19}
\end{eqnarray}
Similarly, the wave function becomes
\begin{equation}
\Psi_{J}=P_{J}(4)\exp\left(-\frac{\vert Z_{1}\vert^{2}}{8l_{0}^{2}}-\frac{\vert Z_{2}\vert^{2}}{6l_{0}^{2}}-\frac{3\vert Z_{3}\vert^{2}}{16l_{0}^{2}}-\frac{\vert Z_{4}\vert^{2}}{l_{0}^{2}}\right).
\label{E20}
\end{equation}
It is possible to develop (\ref{E19}) in terms of $Z_{1}^{n}$, where $n$ belongs to the set
$ \lbrace 3, 5, \ldots, 15 \rbrace $ for the case of $N_{\text{e}}=4$ electrons, thus we have
\begin{equation}
P_{J}(4) = \sum_{n=3}^{15} \mathcal{C}_{n}(Z_{2},Z_{3})\,Z_{1}^{n},
\label{E21}
\end{equation}
where $\mathcal{C}_{n}$ are functions of the only variables $Z_{2}$ and $Z_{3}$ that can be extracted from (\ref{E19}) by the use of
\begin{equation}
\mathcal{C}_{n}(Z_{2},Z_{3})=\frac{1}{\pi\,\Gamma(1+n)}\sum_{m=3}^{15} \mathcal{C}_{m}(Z_{2},Z_{3})\,\int\,\rd^{2}Z_{1}\,Z_{1}^{m}\bar{Z_{1}}^{n}\: e^{-Z_{1} \bar{Z}_{1}},
\label{22}
\end{equation}
wherein the integration is determined with the help of the key rule~\cite{R25}
\begin{equation}
\int \rd^{2}Z\: Z^{m} \,\bar{Z}\,^{n}\: e^{-Z \bar{Z}}=\pi\: \delta_{mn}\:\Gamma(1+n).
\label{E23}
\end{equation}
In computing $\mid \Psi(Z_{1},\ldots ,Z_{3})\mid^{2}$, we will encounter the expression $P_{J}(4) \bar{P_{J}}(4)$, where $\bar{P_{J}}$ is the complex conjugate of $P_{J}$, by requiring to satisfy the rule (\ref{E23}), there only remain the terms with the same power in $Z_{i}$ and $\bar{Z}_{i}$. Thus, the integrand of equation (\ref{E15}) has no dependence on the angles. This independence upon the angles is the key tool that greatly facilitates the exact calculation of complicated expressions involving integrals over many variables (see the work~\cite{R21}). This is the most prominent advantage of the method of complex coordinates. For instance, the integral (\ref{E15}) can be reduced to a simple form
\begin{equation}
\langle\Psi\vert V_{\text{ee}}\vert\Psi \rangle  =  e_{0}^{2}\frac{N_{\text{e}}(N_{\text{e}}-1)}{2}\sum_{n=3}^{15}\mathcal{F}_{n}\int \rd^{2}Z_{1}(Z_{1}\bar{Z}_{1})^{n-\frac{1}{2}}\exp\left(-\frac{\vert Z_{1}\vert^{2}}{4l_{0}^{2}}\right)
\label{E24}
\end{equation}
with the factor $\mathcal{F}_{n}$ given by
\begin{equation}
\mathcal{F}_{n}=\int \rd^{2}Z_{2}\rd^{2}Z_{3}\rd^{2}Z_{4} \mathcal{C}_{n}\bar{\mathcal{C}}_{n}\exp\left(-\frac{\vert Z_{2}\vert^{2}}{3l_{0}^{2}}-\frac{3\vert Z_{3}\vert^{2}}{8l_{0}^{2}}-\frac{2\vert Z_{4}\vert^{2}}{l_{0}^{2}}\right).
\label{E25}
\end{equation}
Similarly, the norm $\langle \Psi\vert\Psi\rangle$ is of the form
\begin{equation}
\langle\Psi\vert\Psi \rangle  =  \sum_{n=3}^{15}\mathcal{F}_{n}\int \rd^{2}Z_{1}(Z_{1}\bar{Z}_{1})^{n}\exp\left(-\frac{\vert Z_{1}\vert^{2}}{4l_{0}^{2}}\right).
\label{E26}
\end{equation}
Now, dividing (\ref{E24}) by (\ref{E26}), we get the (e-e) interaction energy for a system with $N_{\text{e}}=4$ electrons,
$$
E_{\text{ee}}= 1.310596 (e_{0}^{2} / l_{0}),
$$
which coincides with an analogous result in~\cite{R22}.
\section{The $\langle V_{\text{eb}}\rangle$ calculation}
\label{sec:3}
In the case of $\langle V_{\text{eb}}\rangle$ calculation, there is no need to use Jacobi coordinates, it suffices to work with the polynomial $P(4)$ of equation (\ref{E15}) directly. So, let us expand $P(4)$ in powers of $z_{1}$, that is
\begin{equation}
P(4) = \sum_{n=0}^{9} \mathcal{C}_{n}(z_{2}, z_{3}, z_{4})\,z_{1}^{n},
\label{E27}
\end{equation}
where $n = 0$ ($n = 9$) denotes the minimum (maximum) power in $z_{1}$, the wave function is, therefore, written as follows:
\begin{equation}
\Psi=P(4)\exp\left(-\frac{\vert z_{1}\vert^{2}}{2l_{0}^{2}}-\frac{\vert z_{2}\vert^{2}}{2l_{0}^{2}}-\frac{\vert z_{3}\vert^{2}}{2l_{0}^{2}}-\frac{\vert z_{4}\vert^{2}}{2l_{0}^{2}}\right).
\label{E28}
\end{equation}
Furthermore, it is possible to write $\langle\Psi\vert V_{\text{eb}}\vert\Psi \rangle$ like
\begin{equation}
\langle\Psi\vert V_{\text{eb}}\vert\Psi \rangle = \frac{-2N_{\text{e}}^{2}}{R_{\text{e}}} \sum_{n=0}^{9} \mathcal{G}(n)\,\int \mathcal{D}_{p}z\,\mathcal{C}_{n}\bar{\mathcal{C}}_{n}\exp\left(-\frac{\vert z_{2}\vert^{2}}{2l_{0}^{2}} - \frac{\vert z_{3}\vert^{2}}{2l_{0}^{2}} - \frac{\vert z_{4}\vert^{2}}{2l_{0}^{2}}\right)
\label{E29}
\end{equation}
with $\mathcal{D}_{p}z = \rd^{2}z_{2}\rd^{2}z_{3}\rd^{2}z_{4}$,
\begin{equation}
\mathcal{G}(n) = \int \rd r_{1}\int \frac{\rd q}{q}J_{1}(q)J_{0}\left(\frac{q}{R_{N_{\text{e}}}}r_{1}\right)r_{1}^{2n + 1}\exp\left(-\frac{r_{1}^{2}}{2l_{0}^{2}}\right),
\label{E30}
\end{equation}
and $ \vert z_{i}\vert=r_{i} $. One can verify that\,\cite{R26}
\begin{equation}
\mathcal{G}(n) = \left(2l_{0}^{2}\right)^{n + 1}\left(\frac{1}{4}\right)\textnormal{MeijerG} \big[ \big\lbrace \lbrace 1 \rbrace, \lbrace 1 \rbrace \big\rbrace, \big\lbrace \lbrace 1/2, n + 1 \rbrace, \lbrace - 1/2 \rbrace \big\rbrace, N_{\text{e}}m \big].
\label{E31}
\end{equation}
In the present demonstrative calculation, $N_{\text{e}}=4$ and $m=3$. MeijerG is the Meijer G function~\cite{R26}. The expression of the Meijer G function in (\ref{E31}) can also be written as
$$
\textnormal{MeijerG} \big[ \big\lbrace \lbrace 1 \rbrace, \lbrace 1 \rbrace \big\rbrace, \big\lbrace \lbrace 1/2, n + 1 \rbrace, \lbrace - 1/2 \rbrace \big\rbrace, N_{\text{e}}m \big]=G^{21}_{23}\left( N_{\text{e}}\,m \Big\vert^{1,1}_{\frac{1}{2}, n+1, - \frac{1}{2}}\right).
$$
As concerns the norm $\langle\Psi\vert\Psi\rangle$, in the $z$-coordinates, it will take the following form
\begin{eqnarray}
\langle\Psi\vert\Psi \rangle  = \sum_{n=0}^{9}\int \mathcal{D}_{p}z\,\mathcal{C}_{n}\bar{\mathcal{C}}_{n}\exp\left(-\frac{\vert z_{2}\vert^{2}}{2l_{0}^{2}} - \frac{\vert z_{3}\vert^{2}}{2l_{0}^{2}} - \frac{\vert z_{4}\vert^{2}}{2l_{0}^{2}}\right)\int \rd^{2}z_{1}(z_{1}\bar{z}_{1})^{n}\exp\left(-\frac{\vert z_{1}\vert^{2}}{2l_{0}^{2}}\right).
\label{E32}
\end{eqnarray}
As in the case of $V_{\text{ee}}$ calculation, the electron-background interaction energy is determined using
\begin{equation}
E_{\text{eb}}= \frac{\langle\Psi\vert V_{\text{eb}}\vert\Psi \rangle}{\langle\Psi\vert\Psi \rangle}\,,
\label{E33}
\end{equation}
which gives the value $E_{\text{eb}}=-5.638272 (e_{0}^{2}/l_{0})$ as in the work~\cite{R22}. The dependence in angles is simplified when dividing the quantity $\langle\Psi\vert V_{\text{eb}}\vert \Psi\rangle$ by $\langle\Psi\vert \Psi\rangle$. This is because the result of the integration upon the angle variables is the same for both quantities and is equal to $(2\pi)^{4}$.
\section{Results and discussion}
\label{sec:4}
In this paragraph, we present our results concerning the ground state energy for systems with up to $N_{\text{e}} = 7$ electrons at the filling $\nu = 1/3$ and $N_{\text{e}} = 6$ electrons at the filling $\nu=1/5$. We also make a comparison with other works such as~\cite{R22,R17,R27}. Our analysis is summarized in the tables blow. It should be notified that for $N_{\text{e}} = 4$ electrons, the authors in~\cite{R17} and with Monte Carlo calculations, have derived for the energy of the ground state the value $-1.55536\,e_{0}^{2}/l_{0}$ at the filling $\nu = 1/3$ and the value $-1.28636\,e_{0}^{2}/l_{0}$ at the filling $\nu = 1/5$,  which well agree with the values we derived by the exact analytical calculation, see table~\ref{Table1} and table~\ref{Table2}. Table~\ref{Table3} allows us to compare our results concerning the (e-b) and (e-e) interaction energies with those derived in~\cite{R27} at the filling $\nu=1/5$.

\begin{table}[!b]
\caption{\label{Table1} Ground-state energy $E=E_{\text{ee}}+E_{\text{eb}}+E_{\text{bb}}$ (in units of $e_{0}^{2}/l_{0}$) obtained in the Laughlin state at the filling $\nu=1/3$ for systems with up to $N_{\text{e}}= 7$ electrons. The results $E_{\text{A}}$ are the values of the ground state energy obtained also by an exact analytical calculation in~\cite{R22} at the filling $\nu = 1/3$.}
\vspace{2ex}
\begin{center}
\begin{tabular}{|l|l|l|l|l|l|}
  \hline
  \hline
  $N_{\text{e}}$ & $ E_{\text{bb}}$ & $ E_{\text{eb}}$ & $ E_{\text{ee}}$ & $ E$ & $ E_{\text{A}}$ \\
  \hline\hline
   2 & 0.98014  & $-$2.02115   & 0.276946 & $-$0.764064 & $-$0.764064 \\\hline
   3 & 1.800633 & $-$3.679464  & 0.719316 & $-$1.159515 & $-$1.159515 \\\hline
   4 & 2.772256 & $-$5.638272  & 1.310596 & $-$1.55542  & $-$1.55542 \\\hline
   5 & 3.874345 & $-$7.856335  & 2.030715 & $-$1.951275 & --------- \\\hline
   6 & 5.092956 & $-$10.306452 & 2.864394 & $-$2.349102 & --------- \\\hline
   7 & 6.417859 & $-$12.968494 & 3.802267 & $-$2.748368 & ---------  \\
   \hline\hline
\end{tabular}
\end{center}
\end{table}
\begin{table}[!b]
\caption{\label{Table2} Ground-state energy $E=E_{\text{ee}}+E_{\text{eb}}+E_{\text{bb}}$ (in units of $e_{0}^{2}/l_{0}$) obtained in the Laughlin state at the filling $\nu=1/5$ for systems with up to $N_{\text{e}}= 6$ electrons. The results $E_{\text{A}}$ are the values of the ground state energy obtained also by an exact analytical calculation in~\cite{R27} at the filling $\nu = 1/5$.}
\vspace{2ex}
\begin{center}
\begin{tabular}{|l|l|l|l|l|l|}
  \hline
  \hline
  $N_{\text{e}}$ & $ E_{\text{bb}}$ & $ E_{\text{eb}}$ & $ E_{\text{ee}}$ & $ E$ & $ E_{\text{A}}$ \\
  \hline\hline
   3 & 1.394763 & $-$2.911356 & 0.554745 & $-$0.961848 & $-$0.961848 \\\hline
   4 & 2.14738  & $-$4.4429   & 1.009184 & $-$1.286336 & $-$1.286312\\\hline
   5 & 3.001055 & $-$6.17325  & 1.566615 & $-$1.60558  & --------- \\\hline
   6 & 3.944988 & $-$8.082144 & 2.209278 & $-$1.927878 & --------- \\
   \hline\hline
\end{tabular}
\end{center}
\end{table}
\begin{table}[!b]
\caption{\label{Table3} The electron-electron (e-e) and electron-background (e-b) interaction energies are given (in units of $e_{0}^{2}/l_{0}$) at the filling $\nu = 1/5$, namely $E_{\text{ee}}$ and  $E_{\text{eb}}$. Analogous results that we designate by $E_{\text{eeA}}$ and $E_{\text{ebA}}$, are derived in~\cite{R27} at the filling $\nu = 1/5$.}
\vspace{2ex}
\begin{center}
\begin{tabular}{|l|l|l|l|l|}
  \hline
  \hline
  $N_{\text{e}}$ & $ E_{\text{eb}}$ & $ E_{\text{ebA}}$ & $ E_{\text{ee}}$ & $ E_{\text{eeA}}$ \\
  \hline\hline
   3 & $-$2.911356 & $-$2.911356 & 0.554745 & 0.554745 \\\hline
   4 & $-$4.4429   & $-$4.442876 & 1.009184 & 1.009184 \\\hline
   5 & $-$6.17325  & --------- & 1.566615 & -------- \\\hline
   6 & $-$8.082144 & --------- & 2.209278 & -------- \\
   \hline\hline
\end{tabular}
\end{center}
\end{table}

\begin{figure}[!t]
\begin{center}
\includegraphics[height=6cm,width=10cm]{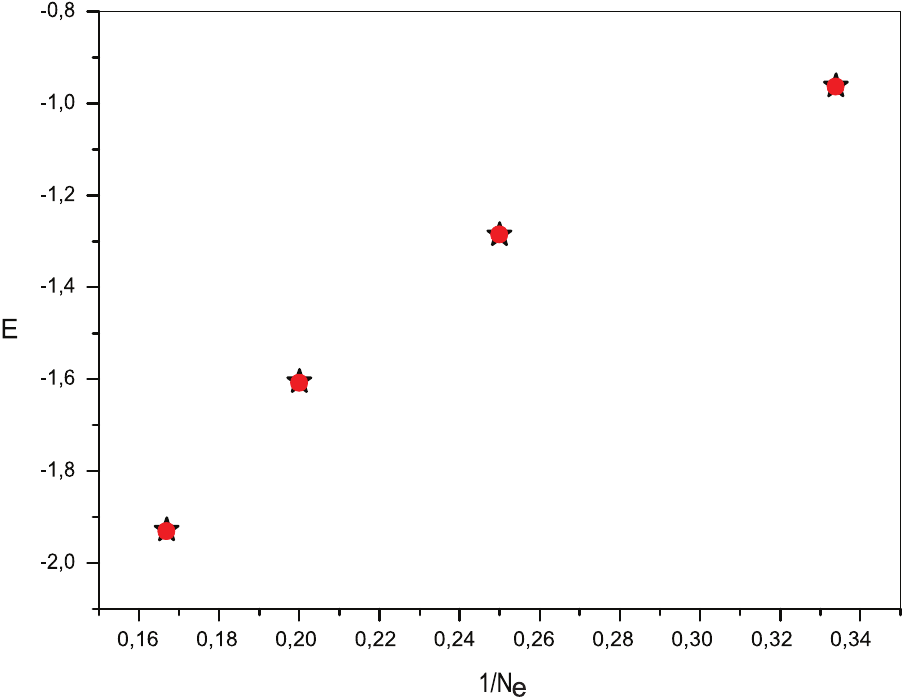}
\end{center}
\caption{\label{fig} (Color online) Exact analytical results for the ground state energy $E$ using the method of complex polar coordinates in disk geometry for the Laughlin state at $\nu=1/5$. The ground state energy $E$ is plotted as a function of $1/N_{\text{e}}$ for systems with $N_{\text{e}}=3,4,5$ and $6$ electrons. The stars represent our result, the disks are the results derived in~\cite{R17} using the method of Monte Carlo. Energies are in units of $e_{0}^{2}/l_{0}$.}
\end{figure}

In tables~\ref{Table1} and \ref{Table2}, our results regarding the energy of the ground state are given in the  fifth column  whereas in the sixth column there are given those of~\cite{R22} and \cite{R27}, respectively. At this point, it should be emphasized that in table~\ref{Table1}, table~\ref{Table2} and table~\ref{Table3}, the comparison is carried out between analytical methods depending on whether ordinary or complex polar coordinates are used. Moreover, this presentation of tables allows one to clearly show the advantage of using polar coordinates in the complex form.

In figure~\ref{fig}, we can see that the results derived by the present exact analytical calculation at the filling $\nu=1/5$ compare well with the results of~\cite{R17} obtained using the method of Monte Carlo.

\section{Concluding remarks}
\label{sec:6}
In this work we have exposed all the necessary steps that permit to make an analytic computation of the energies of the ground state for FQHE systems of electrons at $\nu = 1/m$, $m$ odd. The electron-electron and electron-background interaction energies are calculated separately. The results we derived are in perfect accordance with previous calculations such as the exact analytical calculation of~\cite{R22,R27} or Monte Carlo simulations of~\cite{R17}. In a broader view, the method of complex polar coordinates described in~\cite{R21} may be useful and efficient in analytically calculating the ground state or excited state energies for various quantum Hall systems of electrons with filling factors other than $\nu=1/m$, $m$ odd, such as $(\nu=2/5, 3/7,\ldots )$. We expect that the method of complex polar coordinates has some relevance to 2D~Coulomb systems. For instance, it can be seen, for 2D Dyson gas, that the method of complex coordinates may be useful and practical in analytically evaluating, with many simplifications, the key quantities such as the partition function or the mean energy. The issue of finding links for the approach described in~\cite{R21} with other areas of condensed matter physics remains to be extensively investigated. The calculation can be extended to larger systems with $N_{\text{e}} > 7$ electrons depending on the performance of the machine. This will make it possible to derive exact analytical bulk regime values for key quantities, such as various interaction energies. A part of the code~\cite{R28} of the electron-electron interaction energy computation $V_{\text{ee}}$ is given in the Appendix.

\newpage
\section*{Appendix}
\appendix
{\small \topsep 0.6ex
\begin{verbatim}
* Part of the code of the Vee calculation in MATHEMATICA SOFTWARE (Ne=4 electrons)*

PartialVee=Block[{PolyJaco, CoefPoly, CoefPolyMin, PolyExpand, RePoly, CoefPolyList, Clist,
Inner1, Inlist,  I1, I2, I3,Vee },
PolyJaco=Z[1]^3(-Z[1]/2 + Z[2])^3(Z[1]/2 + Z[2])^3(-2 Z[2]/3 + Z[3])^3
         (-Z[1]/2 + Z[2]/2 + Z[3])^3(Z[1]/2 + Z[2]/2 + Z[3])^3;
CoefPoly=Exponent[PolyJaco, Z[1]];
CoefPolyMin=Exponent[PolyJaco,Z[1],Min];
PolyExpand=Expand[PolyJaco];
RePoly=Flatten[Table[Coefficient[PolyExpand,Z[1],i],{i,CoefPolyMin,CoefPoly,2}]];
CoefPolyList=Plus@@RePoly;
Clist=CoefPolyList/.Plus->List;
Inner1=Inner[Times,Clist/.{Z[2]-> 1, Z[3]-> 1},Clist,Plus];
Inlist=Inner1/.Plus->List/.{Z[2]-> r[2]^2, Z[3]-> r[3]^2};
I1 = Integrate[Inlist*r[2]*r[3]*Exp[-(3 r[3]^2/(8l[0]^2))],
        {r[3],0,Infinity}, Assumptions -> (l[0]>0)];
I2 = Integrate[I1*Exp[-(r[2]^2/(3l[0]^2))],{r[2],0,Infinity},
        Assumptions -> (l[0]>0)];
I3 = Integrate[I2*r[4]*Exp[-(2 r[4]^2/(l[0]^2))],{r[4],0,Infinity},
        Assumptions -> (l[0]>0)];
Vee = Reverse[Plus @@ I3 /. Plus -> List]]]
PartialVee
{1688579923968000 l[0]^36, 3170189352960 l[0]^32,
3024980640 l[0]28, 2095200 l[0]^24, 122715/128 l[0]^20,
189/1024 l[0]^16), 3/262144 l[0]^12}
II[i_] := Integrate[r[1]^(2i)*Exp[(-r[1]^2)/(4l[0]^2)], {r[1], 0,
            Infinity}, Assumptions -> l[0] > 0)];
JJ[i_] := Integrate[r[1]^(2i+1)*Exp[(-r[1]^2)/(4l[0]^2)], {r[1], 0,
            Infinity}, Assumptions -> l[0] > 0)];
list1 = Flatten[Table[II[i], {i, 3, 15, 2}]];
list2 = Flatten[Table[JJ[i], {i, 3, 15, 2}]];
num = Inner[Times, PartialVee, list1, Plus];
denom = Inner[Times, PartialVee, list2, Plus];
FullVee = Divide[num, denom];
Eee = N[Times[FullVee, (12/2) e[0]^2], 6];
Eee
1.310596 e[0]^2 / l[0]
\end{verbatim}
}

\ukrainianpart

\title[]%
{Метод розрахунку енергій у  режимі дробового квантового ефекта Холла}
\author[]{M.A.~Аммар\refaddr{label1},
        З. Бенталха\refaddr{label2},
        С. ~Бехечі\refaddr{label2} }
\addresses{
\addr{label1} Відділення охорони довкілля, Університет м. Медеа, 26000 Медеа, Алжир
\addr{label2} Лабораторія теоретичної фізики, Університет м. Тлемсен, B.P. 230, 13000 Тлемсен, Алжир
}

\makeukrtitle

\begin{abstract}
У попередній роботі ми отримали точні результати для енергій основного стану у  режимі дробового квантового ефекта Холла  (FQHE) для систем з $N_{\text{e}} = 6$ електронів включно при коефіцієнті заповнення  $\nu = 1/3$, використавши метод комплексних полярних координат. В цій роботі ми представляємо цікаві обчислювальні деталі попередніх розрахунків  і розширюємо наші обчислення до  $N_{\text{e}} = 7$ електронів при $\nu = 1/3$.
Крім того, отримано подібні точні результати при заповненні $\nu = 1/5$ для систем з $N_{\text{e}} = 6$ електронів включно. Отримані результати за допомогою аналітичних обчислювань добре узгоджуються з їхніми аналогами, отриманими методом Монте Карло в даній роботі.
\keywords квантовий ефект Холла, 2D електронний газ, багаточастинкова хвильова функція, сильно скорельована система
\end{abstract}

\end{document}